\let\SavedKernelLabel\label
\documentclass[aps,10pt,twocolumn,amsmath,amssymb,superscriptaddress,prx,floatfix]{revtex4-2}
\usepackage{graphicx}
\usepackage{bm}
\usepackage{booktabs}
\usepackage{xcolor}
\usepackage{siunitx}
\let\label\SavedKernelLabel
\makeatletter\let\ltx@label\label\makeatother
\usepackage[colorlinks=true,linkcolor=blue!50!black,citecolor=blue!50!black,urlcolor=blue!50!black]{hyperref}
\hypersetup{pdftitle={Predicting energy and structural response to force correction in molecular dynamics}}
\newcommand{\dd}{\mathrm{d}}
\newcommand{\E}{\mathbb{E}}
\newcommand{\kb}{k_{\mathrm B}}
\newcommand{\rref}{\mathrm r}
\newcommand{\bbase}{\mathrm b}
\newcommand{\eps}{\varepsilon}

\begin{document}
\raggedbottom
\title{Predicting energy and structural response to force correction in molecular dynamics}
\author{Peng Kang}
\affiliation{School of Materials Science and Engineering, Beihang University, Beijing 100191, China}
\affiliation{National Key Laboratory of Artificial Intelligence for Material Science, Beihang University, Beijing 100191, China}
\affiliation{Tianmushan Laboratory, Beihang University, Hangzhou, China}
\author{Da Wan}
\affiliation{School of Materials Science and Engineering, Beihang University, Beijing 100191, China}
\affiliation{National Key Laboratory of Artificial Intelligence for Material Science, Beihang University, Beijing 100191, China}
\affiliation{Tianmushan Laboratory, Beihang University, Hangzhou, China}
\author{Shulin Bai}
\affiliation{School of Materials Science and Engineering, Beihang University, Beijing 100191, China}
\affiliation{National Key Laboratory of Artificial Intelligence for Material Science, Beihang University, Beijing 100191, China}
\affiliation{Tianmushan Laboratory, Beihang University, Hangzhou, China}
\author{Pengfei Zhang}
\affiliation{School of Materials Science and Engineering, Beihang University, Beijing 100191, China}
\affiliation{National Key Laboratory of Artificial Intelligence for Material Science, Beihang University, Beijing 100191, China}
\author{Peng Wang}
\affiliation{School of Materials Science and Engineering, Beihang University, Beijing 100191, China}
\affiliation{National Key Laboratory of Artificial Intelligence for Material Science, Beihang University, Beijing 100191, China}
\author{Chenglong Wen}
\affiliation{School of Materials Science and Engineering, Beihang University, Beijing 100191, China}
\affiliation{National Key Laboratory of Artificial Intelligence for Material Science, Beihang University, Beijing 100191, China}
\author{Zhen Li}
\affiliation{School of Materials Science and Engineering, Beihang University, Beijing 100191, China}
\affiliation{National Key Laboratory of Artificial Intelligence for Material Science, Beihang University, Beijing 100191, China}
\affiliation{Tianmushan Laboratory, Beihang University, Hangzhou, China}
\author{Yu Liu}
\affiliation{School of Materials Science and Engineering, Beihang University, Beijing 100191, China}
\affiliation{National Key Laboratory of Artificial Intelligence for Material Science, Beihang University, Beijing 100191, China}
\affiliation{Tianmushan Laboratory, Beihang University, Hangzhou, China}
\author{Lei Zheng}
\email{zhenglei@buaa.edu.cn}
\affiliation{School of Materials Science and Engineering, Beihang University, Beijing 100191, China}
\affiliation{National Key Laboratory of Artificial Intelligence for Material Science, Beihang University, Beijing 100191, China}
\affiliation{Tianmushan Laboratory, Beihang University, Hangzhou, China}
\author{Li-Dong Zhao}
\email{zhaolidong@buaa.edu.cn}
\affiliation{School of Materials Science and Engineering, Beihang University, Beijing 100191, China}
\affiliation{National Key Laboratory of Artificial Intelligence for Material Science, Beihang University, Beijing 100191, China}
\affiliation{Tianmushan Laboratory, Beihang University, Hangzhou, China}

\date{September 13, 2026}
\begin{abstract}
We predict how force correction changes energy exchange and structural statistics by measuring leading response coefficients on shared reference trajectories. Residual power and the displacement virial distinguish the transfer of energy from the change in restoring forces, including intermittent reference updates. Independent simulations then test the predicted kinetic and configurational shifts. A matched-timetable experiment shows that reference timing affects heating through its coupling to the evolving state. In an anharmonic chain, the displacement virial predicts a structural shift missed by a power-only description. Local force constants in silicon predict a complementary directional tradeoff: scalar calibration repairs optical motion while degrading an already accurate low-frequency direction. Full nonlinear trajectories confirm this tradeoff and distinguish the benefits of static curvature correction and repeated reference impulses. Independent finite-temperature integrals in orthorhombic tin selenide (SnSe) support the configurational-response direction predicted from separate reference calculations. These results provide a physical basis for choosing how reference information enters molecular dynamics. The framework assesses force correction through its effect on atomic motion and statistical observables, beyond the accuracy of individual force evaluations.
\end{abstract}

\maketitle
\section{Introduction}
Predicting the evolution of matter requires following atomic motion over the scales on which collective behavior emerges. Vibrations redistribute energy, local rearrangements enable diffusion, and coupled structural changes govern phase transformations and chemical reactions. Molecular dynamics connects these processes to interatomic forces, but its reach depends on the cost and accuracy of evaluating them. Ab initio molecular dynamics obtains forces from the electronic structure as the system evolves~\cite{car1985}. Machine-learned interatomic potentials extend the accessible length and time scales by representing reference potential-energy surfaces at much lower cost~\cite{behler2007,zhang2018,mace2022}. Pretrained models broaden this opportunity across material chemistries~\cite{macefoundation2025}. The physical value of that speed depends on how force approximations affect the motion and statistical observables for which the simulation is performed.

Reference calculations can improve a simulation as it proceeds. On-the-fly learning acquires electronic-structure data where the potential needs refinement~\cite{csanyi2004,jinnouchi2019}. Multiple-time-step methods combine inexpensive frequent forces with more expensive corrections at longer intervals~\cite{tuckerman1992,barth1998,sandu1999}. State-dependent time stepping can alter the sampled measure, and stability alone need not ensure accurate thermal statistics~\cite{leroy2024,franklin2005}. Harmonic force correction uses local reference force constants to repair a model's restoring forces~\cite{rohskopf2020}. These approaches offer different ways to spend a limited reference budget, and their performance depends on more than the error of an isolated force evaluation. Trajectory benchmarks show that force accuracy alone does not determine the quality of molecular dynamics~\cite{fu2023}. Conservative force construction and smooth potential-energy surfaces are important for reliable propagation~\cite{bigi2025,fu2025smooth}; mode-dependent errors, including systematic phonon softening, affect the predicted material response~\cite{deng2025}. Choosing a correction therefore requires understanding both the force discrepancy and how the correction enters the dynamics.

Our previous work connected a local force-error tolerance to the signed work performed along atomic motion~\cite{kang2026}. Over many reference intervals, this work changes the exchange of energy with the thermostat. Structural statistics also depend on the restoring forces: a conservative force error leaves the canonical momentum distribution unchanged while shifting the configurational distribution. Intermittent reference updates add a coupling between the correction and the trajectory that selects it. Even the full sequence of reference intervals does not specify this feedback. A useful prediction must therefore distinguish energy transfer, restoring-force bias, and the association of reference times with the evolving state. Standard energy and virial balances identify the corresponding force projections; reference-process response estimates~\cite{kubo1957,oden2015} allow them to be evaluated before the corrected dynamics are run.

Here we develop an energy and displacement-virial description of force correction, including held forces and instantaneous impulses. The central step is to evaluate leading response coefficients for candidate corrections on shared reference trajectories, then test their predictions on independent simulations. A matched-timetable experiment isolates the effect of coupling reference times to the evolving state. An anharmonic chain then demonstrates a structural shift missed by a power-only description and tests prospective selection among correction rules. Local force constants in silicon predict a directional tradeoff: a scalar calibration repairs optical motion while spoiling an already accurate low-frequency direction. Full nonlinear trajectories confirm this tradeoff and compare the scalar correction with static-Hessian and symmetric impulse corrections. An independent finite-temperature test in orthorhombic tin selenide (SnSe) then connects the reference coefficient to a nonlinear crystal-response integral. The resulting connection between force error, reference timing, and resolved atomic motion provides a physical basis for allocating expensive reference calculations to a specified dynamical objective.

\section{Energy and displacement-virial response}
\label{sec:response}
A reference-force correction changes the dynamics through both its force and its timing. We describe these effects relative to a specified reference potential $U_{\rref}(\bm q)$. Positions $\bm q$ are measured in a fixed frame, after any fixed linear constraints have been eliminated; $\bm p$ is the conjugate momentum, $\bm M$ the mass matrix, and $\bm v=\bm M^{-1}\bm p$. Between reference impulses, the approximate Langevin dynamics obey
\begin{align}
 \dd\bm q &= \bm v\,\dd t,\nonumber\\
 \dd\bm p &= [-\nabla U_{\rref}+\bm R-\gamma\bm p]\,\dd t
       +\sqrt{2\gamma\tau\bm M}\,\dd\bm W,
 \label{eq:langevin}
\end{align}
where $\bm R$ is the approximate-minus-reference continuous force, $\tau=\kb T_{\mathrm b}$, and the friction $\gamma$ is common to the $d$ unconstrained degrees of freedom. An impulse at time $t_k$ changes momentum by $\Delta\bm p_k$ without changing position. The dynamics may include a stored reference configuration and an update clock.

\begin{figure*}[!tp]
\centering
\includegraphics[width=\textwidth]{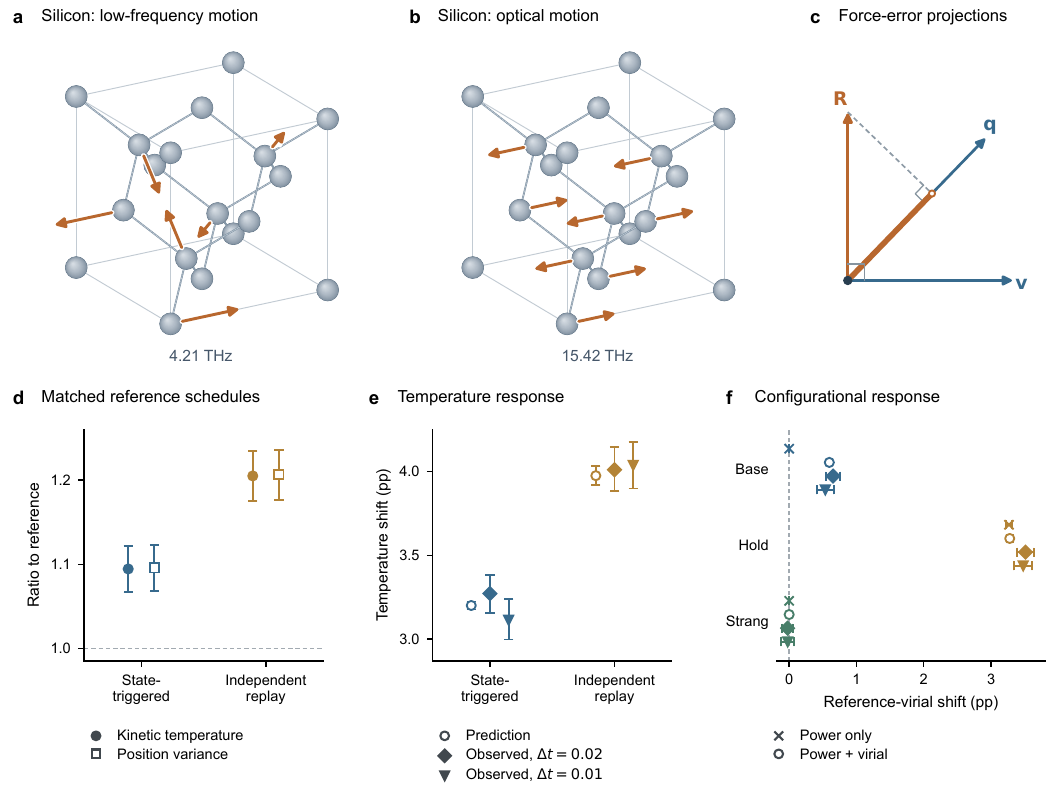}
\caption{\label{fig:overview}\textbf{Force-error projections and their statistical consequences.} (a,b) Low-frequency and optical displacement directions in the eight-atom silicon cell, from the 4.21\,THz sixfold and 15.42\,THz threefold reference subspaces. Spheres mark reference positions; thin lines show nearest-neighbor bonds and the cell boundary. Boundary images complete the periodic cell. Copper arrows mark nonzero displacements of the original eight atoms, with a common viewing direction and linear magnification. They show normalized directions, rather than excitation amplitudes. (c) A residual force $\bm R$ perpendicular to velocity $\bm v$ has a nonzero displacement projection onto $\bm q$ (copper segment); the dashed line constructs that projection. (d) Harmonic oscillator: feedback and independent replay share each pair's complete reference-query timetable, with a mean of 5025.7 evaluations per trajectory including initialization. Initial states and noise are independent, and each trajectory uses its own configuration for the correction. (e) Anharmonic chain: reference-process predictions and a subsequent independent 256-pair temperature validation; the two validation grids have coupled noise. (f) Reference-virial response in the anharmonic chain. Power-only and power-plus-virial predictions use 128 reference paths; observations use 256 new paths per grid, independent between grids and paired across methods. Diamonds and triangles denote the same observation grids as in (e). The base virial resolves a structural shift absent from the leading power; the held response is underpredicted, whereas Strang's intervals include zero. In (e,f), shifts are relative to a same-grid numerical reference and are expressed in percentage points (pp). Fixed, mean-zero thermostat-noise controls reduce variance; observed virials are measured directly from coordinates. All error bars in (d)--(f) are 95\% trajectory- or pair-bootstrap intervals. Statistical runs span 1200 reduced time units, with the first 200 discarded.}
\end{figure*}

The residual force has two distinct projections [Fig.~\ref{fig:overview}(c)]. Its projection onto velocity supplies the residual power,
\begin{align}
 \mathcal P={}&\langle\bm v\cdot\bm R\rangle\nonumber\\
 &+\lim_{T\to\infty}\frac1T\E\sum_{t_k<T}\left[\bm v_k^-\cdot\Delta\bm p_k
 +\frac12\Delta\bm p_k^{\mathsf T}\bm M^{-1}\Delta\bm p_k\right],
 \label{eq:power}
\end{align}
while its projection onto displacement supplies
\begin{equation}
 \mathcal V=\langle\bm q\cdot\bm R\rangle
 +\lim_{T\to\infty}\frac1T\E\sum_{t_k<T}\bm q_k\cdot\Delta\bm p_k.
 \label{eq:residual-virial}
\end{equation}
Angle brackets denote averages over physical time. The quadratic impulse term in Eq.~\eqref{eq:power} is part of the kinetic-energy change and is retained even when the impulse is corrective.

Let $K=\bm p^{\mathsf T}\bm M^{-1}\bm p/2$ and define the reference displacement virial $\Xi_{\rref}=\bm q\cdot\nabla U_{\rref}$. In a stationary, confined process with the required finite moments, the stochastic energy and virial balances give
\begin{align}
 \theta_K-1&=\frac{\mathcal P}{d\gamma\tau},\label{eq:temperature}\\
 \theta_{\Xi}-1&=\frac{\mathcal P}{d\gamma\tau}
                    +\frac{\mathcal V}{d\tau},
 \label{eq:structure}
\end{align}
where $\theta_K=2\langle K\rangle/(d\tau)$ and $\theta_{\Xi}=\langle\Xi_{\rref}\rangle/(d\tau)$. The identities follow by applying the Langevin generator to $K+U_{\rref}$ and $\bm q\cdot\bm p$, including each momentum jump. Appendix~\ref{app:balances} gives the finite-window equations and the conditions for their stationary limits. Fixed-interval updates are averaged over their clock phase.

Equation~\eqref{eq:temperature} relates heating to residual work against thermostat dissipation. Equation~\eqref{eq:structure} adds the restoring-force projection that governs the displacement virial. For a harmonic reference it measures configurational second moments; for an anharmonic reference it includes the corresponding higher-order terms. The displacement virial is used here as a structural observable in confined or translation-reduced coordinates, rather than as a definition of pressure in a periodic material.

\subsection{Prediction from the reference process}
The balances become predictive for a small force perturbation. Write $\bm R_{\eps}=\eps\bm r$ and $\Delta\bm p_k=\eps\bm k_k$. Suppose the required stationary time and event averages are continuous as $\eps\to0$, with a uniformly bounded impulse second-moment rate, and interchangeable long-time and small-perturbation limits. Then
\begin{equation}
 \mathcal P_{\eps}=\eps\mathcal A_0+o(\eps),\qquad
 \mathcal V_{\eps}=\eps\mathcal B_0+o(\eps),
 \label{eq:response-coefficients}
\end{equation}
where
\begin{align}
 \mathcal A_0&=\langle\bm v\cdot\bm r\rangle_0
       +\lim_{T\to\infty}\frac1T\E_0\sum_{t_k<T}\bm v_k\cdot\bm k_k,
 \nonumber\\
 \mathcal B_0&=\langle\bm q\cdot\bm r\rangle_0
       +\lim_{T\to\infty}\frac1T\E_0\sum_{t_k<T}\bm q_k\cdot\bm k_k.
 \label{eq:reference-estimators}
\end{align}
The subscript 0 denotes the reference dynamics, with the candidate anchor and clock rule evaluated along that process. Equations~\eqref{eq:temperature}--\eqref{eq:reference-estimators} predict the leading energy and structural shifts without running each corrected dynamics to stationarity. This is a specialization of response and observable-based error analysis~\cite{kubo1957,oden2015} to force updates whose energy and virial terms can be measured directly.

The power and virial definitions use the residual force directly and apply also to nonconservative force predictions~\cite{bigi2025}. Conservativity gives an additional simplification: the work of a held correction can then be evaluated from segment endpoints. For a general residual, the same response coefficients involve its path integral and event contributions; their evaluation cost depends on the available force information.

For a held reference-force correction, let $U_{\bbase}=U_{\rref}-\eps V$ and let $\bm a$ be the last reference configuration. The applied force is $-\nabla U_{\bbase}(\bm q)-\eps\nabla V(\bm a)$, so
\begin{equation}
 \bm R=\eps[\nabla V(\bm q)-\nabla V(\bm a)].
\end{equation}
Its exact work over a segment ending at $\bm b$ is
\begin{align}
 W_{\bm a\to\bm b}&=\eps D_V(\bm b,\bm a),\nonumber\\
 D_V(\bm b,\bm a)&=V(\bm b)-V(\bm a)
                  -\nabla V(\bm a)\cdot(\bm b-\bm a).
 \label{eq:segment-work}
\end{align}
This Taylor remainder connects local directional work to an accumulated response over many reference intervals. For $\eps\ge0$ and convex $V$, each completed segment contributes nonnegative work. Its mean rate depends on the joint distribution of anchor, displacement, and duration. It is the sum of segment works divided by total elapsed time; averaging segment power with equal weight generally gives a different quantity.

Two familiar correction choices provide useful limits. For the conservative base force alone, $\mathcal A_0=0$, because $\bm v\cdot\nabla V$ is a total time derivative. Its $\mathcal B_0=\langle\bm q\cdot\nabla V\rangle_0$ can remain nonzero. A symmetric Strang correction with a fixed outer interval applies half impulses from $-\eps\nabla V$ around base propagation. With exact fast propagation on the stationary reference process, the mean continuous and endpoint contributions cancel at first order, giving $\mathcal A_0=\mathcal B_0=0$. This identifies the leading response of an established multiple-time-step construction~\cite{tuckerman1992}; the cancellation is in the reference expectation, not segment by segment.

\subsection{A local configurational probe for crystals}
The restoring-force response can be tested in a periodic crystal without assigning a global displacement from one lattice configuration. In a fixed local coordinate chart with translation removed, choose a smooth, compactly supported vector field $\bm h(\bm q)$ and set
\begin{equation}
 X_h=\frac{\bm h\cdot\nabla U_{\rref}}{\tau},\qquad
 D_h=\nabla\cdot\bm h,\qquad
 R_h=\bm h\cdot(\bm F_{\bbase}-\bm F_{\rref}).
\end{equation}
For $U_\lambda=(1-\lambda)U_{\rref}+\lambda U_{\bbase}$, let $I_\lambda[f]=\int f\exp(-U_\lambda/\tau)\dd\bm q$. Integration by parts gives
\begin{equation}
 \theta_{h,\lambda}\equiv\frac{I_\lambda[X_h]}{I_\lambda[D_h]}
 =1+\lambda\frac{I_\lambda[R_h]}{\tau I_\lambda[D_h]}.
 \label{eq:crystal-probe}
\end{equation}
Compact support removes the boundary flux, and the unknown configurational partition function cancels. When the denominator stays nonzero and the integrals are differentiable near $\lambda=0$,
\begin{equation}
 \theta_{h,\lambda}=1+\lambda\beta_{h,0}+o(\lambda),\qquad
 \beta_{h,0}=\frac{I_0[R_h]}{\tau I_0[D_h]}.
 \label{eq:crystal-forecast}
\end{equation}
This extends the conservative displacement-virial test to local crystal fluctuations. It measures the reference restoring-force response in the region selected by $\bm h$. Nonlinear reference energies and forces determine the integrals; harmonic force constants can supply an efficient sampling distribution without replacing that response by a harmonic calculation.

\section{Reference timing and statistical response}
\label{sec:statistics}

\subsection{The same timetable produces different heating}
A reference schedule carries more information than its mean query rate. To isolate its coupling to the trajectory, we first use a harmonic reference $U_{\rref}=q^2/2$ and a softer base potential $U_{\bbase}=0.95q^2/2$, with unit mass and thermal energy, and friction $\gamma=0.05$. The reference-minus-base force at the anchor is held as an additive correction until the displacement from the anchor reaches 0.2 or the anchor age reaches 0.5. For each realization, a second trajectory receives the first trajectory's complete query timetable. It evaluates each reference correction at its own instantaneous position. The two trajectories therefore use exactly the same query counts, intervals, and ordering, while their initial conditions and stochastic forcing are independent. We call them the feedback and replay trajectories, respectively.

The replay changes the thermal response substantially [Fig.~\ref{fig:overview}(d)]. Across 256 independent pairs, the mean squared velocity is 1.0944 for feedback and 1.2050 for replay. Their paired difference is $-0.1107$, with a 95\% confidence interval of $[-0.1532,-0.0685]$. The mean squared displacement changes by almost the same amount: its paired difference is $-0.1108$, with interval $[-0.1531,-0.0688]$. Halving the event-monitoring interval from 0.02 to 0.01 gives differences of $-0.1113$ and $-0.1115$, respectively. Propagation between monitoring points is exact for the held harmonic force. The result therefore resolves the physical effect of associating reference times with the evolving state, rather than an error from integrating the harmonic equations.

In this example the segment work is proportional to the squared displacement from the anchor. Feedback limits that displacement, whereas the replay timetable is generated without feedback from the receiver's state. Matching every interval cannot match the work accumulated within those intervals. This comparison identifies the role of trajectory conditioning at fixed query counts. Replay has 4.25 times the residual second moment of feedback, but only 2.17 times its excess kinetic temperature (Table~\ref{tab:response-checks}). The unequal ratios show that residual magnitude does not give a proportional account of the thermal shift; the energy balance requires its signed projection onto velocity. An independent anharmonic chain test confirms the direction of the effect [Fig.~\ref{fig:overview}(e)]. A prediction made from reference trajectories gives a feedback-minus-replay temperature shift of $-0.775$ percentage points. New validation paths give $-0.739$ and $-0.924$ percentage points on the two integration grids, with both confidence intervals below zero. The paired fine-minus-coarse change in this contrast is $-0.185$ percentage points, with a 95\% interval of $[-0.391,0.027]$; the cross-grid change is unresolved at this sampling precision.

\subsection{A structural response remains when the power vanishes}
The energy and virial terms can be separated in a periodic, eight-particle anharmonic chain. The reference potential is $U_{\rref}=U_2+U_4$, and the base potential is $U_{\bbase}=0.99U_2+U_4$, where $U_2$ and $U_4$ contain quadratic and quartic bond terms. Removing translation leaves $d=7$ degrees of freedom. We set $\tau=\kb T_{\mathrm b}=0.25$ and $\gamma=0.05$. The relevant structural quantity is the \emph{total} reference displacement virial,
\begin{equation}
 \Xi_{\rref}=2U_2+4U_4,
 \label{eq:chain-virial}
\end{equation}
which includes the anharmonic part of the restoring force.

For the uncorrected conservative base potential, the leading residual power is zero. Retaining only the power term in Eq.~\eqref{eq:structure} would predict zero normalized virial shift. The displacement virial supplies the missing structural contribution [Fig.~\ref{fig:overview}(f)]. Using 128 reference calibration paths, it predicts a normalized shift of $0.597\%$ in $\Xi_{\rref}$. Independent validation with 256 paths gives $0.652\%$ at step 0.02 and $0.537\%$ at step 0.01. Their 95\% intervals, $[0.544,0.758]\%$ and $[0.413,0.663]\%$, overlap the prediction, and the corresponding prediction-error intervals contain zero. Temperature-shift intervals include zero on both grids. Thus the reference structure responds even when the energy balance alone gives no leading signal.

The same reference paths predict the response to held and symmetric impulse corrections with an outer interval of 0.2. For the held correction, the predicted virial shift is $3.277\%$, and the observed shifts are $3.512\%$ and $3.476\%$. The leading response captures the scale but underestimates the finite-perturbation shift by about 6\%; the discrepancy is resolved statistically. Table~\ref{tab:response-checks} reports the held-correction prediction errors and the base cross-grid contrast, together with the harmonic feedback and replay residual second moments. For the symmetric Strang correction, both leading coefficients vanish, and the measured virial-shift intervals include zero on both grids. These three cases distinguish a conservative structural bias, simultaneous heating and structural bias under a held correction, and cancellation of the leading response under a symmetric correction.

Confidence intervals here are obtained by resampling independent trajectories. To resolve small shifts efficiently, we subtract analytically centered thermostat-noise terms with coefficients fixed by the balance equations. This variance reduction uses the measured positions, momenta, and thermostat increments; it does not construct the observed response from residual-work predictions. Appendix~\ref{app:statistics} gives the estimator and both raw and variance-reduced results.

\subsection{Selecting a correction with the response prediction}
The leading power can also rank candidate reference schedules before deploying them. For the anharmonic chain, we compare 17 clocks whose next interval is inversely proportional to the speed at the current anchor, with a maximum age of 0.5. All candidates are evaluated on the same 128 reference paths. We choose the clock with the smallest predicted absolute temperature shift among those whose calibration query rate does not exceed five per unit time. This fixes the clock parameter at 0.205 before independent validation.

Table~\ref{tab:selection} compares this choice with a fixed held correction and the standard Strang correction. At approximately matched query counts, the selected clock lowers the held-correction temperature bias by 8.6\% and 5.1\% on the two grids. The paired gains are 0.298 and 0.174 percentage points, respectively, and are resolved on both grids. The finer-grid gain is 42\% smaller. Halving the integration step requantizes the selected clock's intervals at the same fixed parameter, changing its realized query count. The two comparisons therefore test the rule as implemented on each grid. Strang performs substantially better at the same nominal outer interval. The response calculation therefore gives both a modest improvement within a restricted family and a reason to prefer a different correction rule when it is available. The selected parameter is not refitted to the validation paths. A separately specified bond-fluctuation observable, $U_2/d$, improves on the coarser grid; its change on the finer grid is unresolved.

\begin{table}[!tbp]
\caption{\label{tab:selection}Independent temperature validation in the anharmonic chain. Shifts are percentages of the target temperature. Parentheses give 95\% trajectory-bootstrap intervals. Paired gain is the fixed-hold minus selected-clock absolute temperature bias, in percentage points (pp). Query counts cover the full time 1200 and include initialization. Predictions use a separate reference ensemble.}
\centering
\begin{ruledtabular}
\begin{tabular}{lrr}
 & Fixed hold & Selected clock\\
\hline
Predicted shift (\%) & 3.221 & 3.074\\
Observed, $\Delta t=0.02$ & 3.471 & 3.173\\
95\% interval & (3.353,3.587) & (3.050,3.296)\\
Queries, $\Delta t=0.02$ & 6001 & 5998.8\\
Paired gain (pp) & \multicolumn{2}{c}{$0.298\;(0.188,0.414)$}\\
Observed, $\Delta t=0.01$ & 3.375 & 3.201\\
95\% interval & (3.256,3.489) & (3.089,3.315)\\
Queries, $\Delta t=0.01$ & 6001 & 5846.8\\
Paired gain (pp) & \multicolumn{2}{c}{$0.174\;(0.051,0.305)$}\\
\end{tabular}
\end{ruledtabular}
\smallskip
\raggedright\footnotesize
Strang uses 6001 queries and gives $0.025\%$ ($-0.065,0.111$) and $-0.025\%$ ($-0.108,0.062$), respectively. The selected clock's post-equilibration query rate is 5.013 and 4.892; its validation budget is therefore approximately matched, rather than a strict rate bound.
\end{table}

\section{Directional stiffness in silicon}
\label{sec:silicon}
\begin{figure*}[!tp]
\centering
\includegraphics[width=\textwidth]{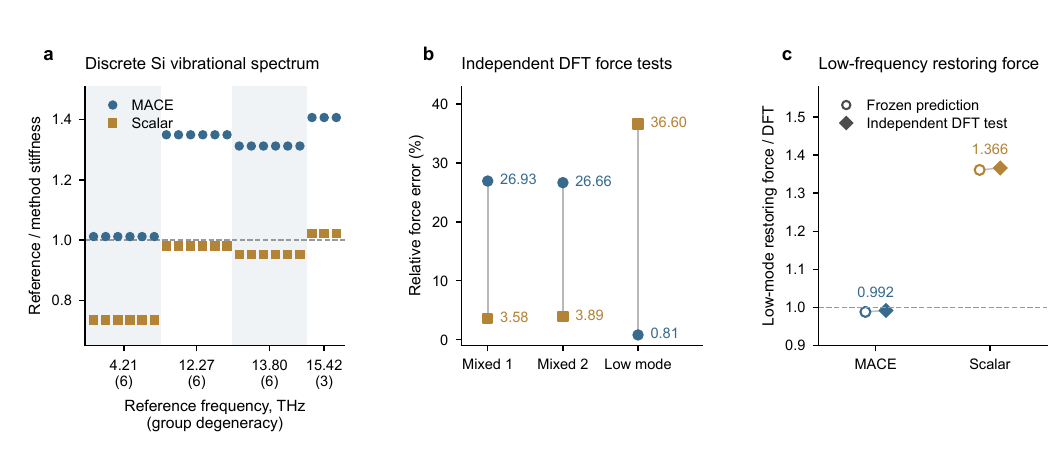}
\caption{\label{fig:silicon}\textbf{Scalar calibration redistributes the directional error in silicon.} (a) Reference-to-method stiffness ratios for the 21 nontranslational modes of the eight-atom cell. The four groups are labeled by reference frequency and degeneracy; they are discrete cell modes, not a Brillouin-zone dispersion. The fixed scalar $c=1.3768451271$ improves higher-frequency stiffnesses and makes the lowest group too stiff. (b) Full Cartesian relative force errors, $100\|\bm F-\bm F_{\mathrm{DFT}}\|_2/\|\bm F_{\mathrm{DFT}}\|_2$, at two mixed-displacement configurations and a preselected low-frequency configuration. Atomic RMS displacements are 0.01, 0.02, and 0.02\,\AA, respectively. Each point is one independent test geometry, and none was used to fit $c$. (c) Low-direction restoring-force ratios predicted before the final DFT evaluation and then measured. The ratios use the corresponding predicted or measured reference force as denominator. Scalar calibration increases the measured ratio from 0.992 to 1.366, confirming the direction of the local prediction.}
\end{figure*}

The chain tests establish the statistical response under controlled conditions. In silicon, local harmonic analysis predicts directional stiffness and covariance errors; separate finite-time nonlinear trajectories test the resulting modal waveforms. We compare a pretrained MACE potential with a fixed density-functional reference, using an eight-atom cubic cell at lattice constant $5.431$\,\AA. The reference is PBE with D3 dispersion. All discrepancies below refer to this specified potential pair; they include any difference between the electronic-structure target and the model's training target.

The local force constants reveal an uneven distribution of relative errors [Fig.~\ref{fig:silicon}(a)]. The 21 nontranslational reference modes form groups at 4.21, 12.27, 13.80, and 15.42\,THz. The corresponding reference-to-base stiffness ratios are 1.012, 1.349, 1.312, and 1.407. The lowest group is already close to the reference, while the higher-frequency groups are softer in the base model. This distribution matters because a global force error and a relative mode error weight the same force-constant discrepancy differently.

A scalar force calibration, fixed at $c=1.37685$ from an earlier optical-direction reference calculation, illustrates the consequence. It brings the higher-frequency stiffness ratios close to unity but lowers the lowest-group ratio to 0.735. On two independent mixed-displacement configurations with atomic root-mean-square displacements of 0.01 and 0.02\,\AA, the relative force error drops from 26.93\% and 26.66\% to 3.58\% and 3.89\%. On a low-frequency displacement specified before its reference evaluation, the same calibration increases the error from 0.81\% to 36.60\%. A correction that improves the larger restoring forces can spoil a direction that was already accurate.

The local spectrum also quantifies the best result available to \emph{any} scalar calibration for harmonic configurational covariance. On the positive force-constant subspace, let
\begin{equation}
 \bm G=\bm H_{\rref}^{1/2}\bm H_{\bbase}^{-1}\bm H_{\rref}^{1/2},
 \qquad a=\lambda_{\min}(\bm G),\quad b=\lambda_{\max}(\bm G).
\end{equation}
Scaling the base force by $c>0$ gives the reference-normalized covariance $\bm G/c$. Elementary minimization then yields
\begin{equation}
 \min_{c>0}\|\bm G/c-\bm I\|_{\mathrm{op}}
 =\frac{b-a}{b+a},\qquad c_* =\frac{a+b}{2}.
 \label{eq:scalar-bound}
\end{equation}
For this potential pair the minimum is 16.33\%. This is a statement about the local harmonic covariance, not a lower bound on a finite-time trajectory error. It expresses the directional information lost by a scalar correction and motivates a comparison with a full force-constant correction. The tradeoff also survives two broader force fits. Minimizing force MSE over isotropic or reference-harmonic thermal displacements gives mixed-configuration force errors of 3.7--4.5\%, while the low-frequency error rises from 0.81\% to 31.5--33.7\% (Table~\ref{tab:scalar-controls}). The directional loss therefore persists when the scalar is chosen using all reference force constants rather than one optical displacement.

\section{Independent nonlinear material dynamics}
\label{sec:dynamics}
\begin{figure*}[!tp]
\centering
\includegraphics[width=\textwidth]{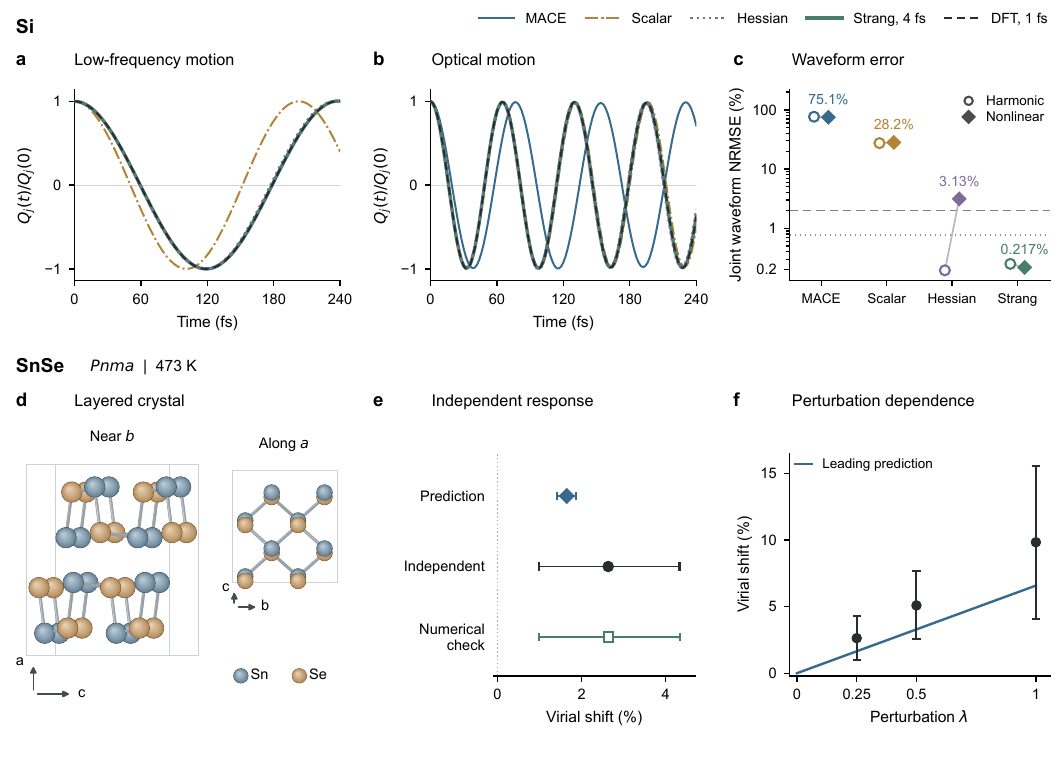}
\caption{\label{fig:materials}\textbf{Complementary material tests of force correction and configurational response.}
(a,b) Silicon low-frequency and optical coordinates divided by their fixed initial amplitudes, 0.226274 and 0.056569\,\AA, over the complete 240\,fs interval. All Cartesian coordinates evolve from the same mixed displacement at rest. Hessian denotes static full-Hessian correction; Strang uses a 4\,fs outer interval.
(c) Harmonic predictions and nonlinear joint waveform errors [Eq.~\eqref{eq:waveform}] on the common 2\,fs grid, without alignment or refitting. The dashed line marks the predetermined 2\% target; the dotted line gives the 0.760\% difference between the two DFT references as a resolution diagnostic. The static-Hessian forecast underestimates the nonlinear error. Deployment requires 241, 121, and 61 reference evaluations for DFT at 1\,fs, DFT at 2\,fs, and Strang, respectively, including initialization; base, scalar, and static-Hessian propagation use zero. Calibration and independent scoring are separate costs.
(d) Tin selenide (SnSe) reference crystal. The side view keeps the stacking axis $a$ vertical and is rotated $12^\circ$ from $b$ toward $c$ to resolve the depth of the puckered layers. It shows a $1\!\times\!2\!\times\!2$ display repeat. The view along $a$ shows one puckered double layer repeated twice along $b$ and $c$. Both are orthographic projections of the unchanged eight-atom reference cell; sticks connect Sn--Se neighbours separated by 2.5--3.2\,\AA.
(e) Primary normalized virial shift $100(\theta_{h,0.25}-1)$, predicted using 32 reference configuration pairs and observed using 64 independent DFT pairs and 16\,384 independent model pairs. Prediction bars propagate calibration uncertainty with $t_{31}$; observation bars are approximate 95\% Fieller intervals. The numerical check replaces the most influential pair and its energy origin with stricter DFT results in a sensitivity copy of the same sample.
(f) All three predefined nonzero perturbations of the primary SnSe probe. The line gives the leading prediction from the frozen reference coefficient. The observations share configurations and are correlated; the controlled zero-perturbation identity is not an independent observation. Appendix~\ref{app:snse} gives the estimator and probe-size sensitivity.}
\end{figure*}

We next test whether the local directional analysis survives full nonlinear dynamics. The initial displacement combines a low-frequency collective direction and an optical direction, with atomic root-mean-square amplitudes of 0.08 and 0.02\,\AA, respectively; all initial velocities are zero. The low-frequency direction is obtained by projecting one atomic Cartesian displacement onto the lowest reference mode subspace. Both directions, amplitudes, and the full 240\,fs comparison window are fixed before evaluating the new trajectories.

Four approximate propagators start from this state: the base MACE force, the scalar-calibrated force, a static full-Hessian correction, and a symmetric Strang correction with a 4\,fs outer interval and a 0.5\,fs inner step. The static correction adds the complete reference-minus-base harmonic restoring force at the equilibrium configuration,
\begin{equation}
 U_{\mathrm H}(\bm q)=U_{\bbase}(\bm q)
 +\tfrac12\bm q^{\mathsf T}(\bm H_{\rref}-\bm H_{\bbase})\bm q.
 \label{eq:hessian-correction}
\end{equation}
This established harmonic force-correction strategy~\cite{rohskopf2020} is a strong comparator: it recovers the entire reference Hessian and uses no additional reference forces during propagation. The Strang method instead applies half impulses from $\bm F_{\rref}-\bm F_{\bbase}$ at the ends of each outer interval~\cite{tuckerman1992}. Each impulse uses the reference force at that trajectory's actual configuration.

We measure both modal waveforms against an independently propagated density-functional trajectory. For mode $j$, the error $E_j$ is the time root-mean-square difference divided by its fixed initial amplitude; $E_{\mathrm{joint}}$ is the root mean square of the two $E_j$ values. The comparison uses the entire 240\,fs interval, with no phase alignment or amplitude fitting. A joint error of 2\% is the criterion chosen before the complete nonlinear trajectory comparison.

The material dynamics retain the directional tradeoff [Fig.~\ref{fig:materials}(a--c)]. Against the 1\,fs reference, the base potential has a 0.60\% low-frequency error but a 106.14\% optical error. Scalar calibration reduces the optical error to 6.23\% while increasing the low-frequency error to 39.41\%. Their joint errors are 75.05\% and 28.22\%, respectively.

The full-Hessian correction reduces the joint error to 3.13\%. It restores the reference curvature at equilibrium, while higher-order restoring forces remain those of the base potential. In the harmonic comparison, its 0.194\% joint error arises from the different propagation steps, 0.5\,fs for the corrected model and 1\,fs for the reference. The nonlinear reference waveform itself differs from its harmonic prediction by 2.50\%, showing a measurable finite-amplitude response. The remaining corrected-model error is consistent with anharmonic force differences beyond the restored local curvature.

Strang gives a joint error of 0.217\% against the 1\,fs reference, and 0.674\% against an independent 2\,fs reference. Both comparisons meet the fixed 2\% criterion; the static-Hessian errors, 3.13\% and 2.49\%, exceed it for both references. Halving the approximate propagators' inner steps changes each waveform by less than 0.05\% in the same joint normalization. The two density-functional references differ by 0.760\%, so the smaller Strang point error should be read together with this resolution check. The result establishes the method ordering and the 2\% classification at the tested resolutions.

The reference-energy peak-to-peak variation is 0.123\% of the initial excitation energy at the 1\,fs step. Strang uses 61 reference evaluations, including initialization, compared with 121 for the 2\,fs reference propagation. This is a reduction in deployment reference evaluations for the specified waveform target. The prior force-constant calibration and the denser validation reference are separate costs, listed in Appendix~\ref{app:silicon}.

\section{Finite-temperature configurational response in tin selenide}
\label{sec:snse}
Thermoelectric materials enable direct heat-to-electricity conversion and solid-state cooling, making control of electronic and thermal transport an important materials-design objective~\cite{bai2026thermoelectrics,zhao2016snse,he2019snssese,liu2023plainification}. Thermal expansion, strain-dependent bonding, higher-order phonon scattering, and interlayer geometry make predicting lattice dynamics and heat transport demanding, with accuracy tied to the interatomic forces~\cite{wan2025mgthermal,wan2025bicu,bai2025rashba,kang2026twistsnse}. We therefore use tin selenide (SnSe), whose low thermal conductivity is linked to strongly anharmonic bonding~\cite{zhao2014snse,xiao2016snseorigin}, to test the predicted finite-temperature configurational response.

We use the eight-atom orthorhombic $Pnma$ cell at 473\,K, with a PBE--D3 density-functional reference and the same pretrained MACE model used for silicon. The fixed cell has lattice parameters $a=11.5524$, $b=4.1777$, and $c=4.4261$\,\AA\ [Fig.~\ref{fig:materials}(d)]. Removing translation leaves 21 displacement coordinates. The probe $\bm h=\bm q\chi(s)$ selects fluctuations around the reference crystal, where $s=\bm q^{\mathsf T}\bm H_{\rref}\bm q/\tau$. The smooth cutoff is one for $s\le20$ and zero for $s\ge40$. Its complete divergence, $D_h=21\chi+2s\chi'$, enters Eq.~\eqref{eq:crystal-probe}.

The prediction was fixed before the independent material calculation. A reference calibration using 32 independent configuration pairs gave $\beta_{h,0}=0.06586$ with standard error 0.00446. At the primary perturbation $\lambda=0.25$, Eq.~\eqref{eq:crystal-forecast} therefore predicts a $1.65\%$ increase in the normalized reference virial. New nonlinear density-functional energies and forces on 64 independent pairs give a $2.63\%$ increase, with an approximate 95\% interval of $[0.99,4.33]\%$ [Fig.~\ref{fig:materials}(e)]. All 128 reference configurations are retained. The observation removes zero-order sampling noise using the exact identity $I_0[X_h-D_h]=0$, retaining the change induced by the perturbation. The reference-level response integrand uses reference forces and both potential energies, without the calibrated coefficient or base forces. Surrogate forces enter the variance-reduction calculation. An independent pool of 16\,384 inexpensive model pairs reduces the integration cost, with both sampling contributions retained in the uncertainty.

The reference calculation thus anticipates the direction of a configurational change in a chemically distinct crystal. Including calibration uncertainty, the observed-minus-predicted shift is 0.99 percentage points, with a 95\% interval of $[-0.73,2.70]$ percentage points. The data support a positive response and are compatible with the leading prediction at this precision. The point estimates rise to $5.09\%$ and $9.84\%$ at the two predefined larger perturbations, $\lambda=0.5$ and 1 [Fig.~\ref{fig:materials}(f)]. The same frozen coefficient sets the prediction at all three perturbations.

The residual-force coefficient from the validation sample agrees with the independent calibration within one combined standard error (Appendix~\ref{app:estimator-equivalence}). The zero-perturbation energy-response derivative, calculated from the observation estimator, has a standard error 6.1 times as large. Shared sampling fluctuations and possible nonlinear response both enter the interpretation of the three observations' departures from the leading prediction.

The same estimation procedure was tested on harmonic and anharmonic models with known nonzero responses. Across 250 independent replicates per model, nominal 95\% intervals covered the true values in 94.8--96.8\% of replicates at each nonzero perturbation (Appendix~\ref{app:known-answer-response}).

The numerical checks distinguish electronic-structure sensitivity from statistical sampling uncertainty. Recomputing the most influential configuration pair with higher plane-wave cutoffs and a denser $k$ mesh changes the primary response by 0.0053 percentage points. This change is small compared with the statistical standard error of 0.83 percentage points. The nonlinear reference--surrogate difference contributes 97.2\% of the estimated response variance. Sampling this difference more efficiently could sharpen the quantitative resolution and extend the comparison to weaker perturbations. Appendix~\ref{app:snse} gives the paired estimators and probe-size and numerical sensitivity checks.

\section{Physical interpretation and outlook}
The useful accuracy of a force model depends on the motion and observable for which it is used. Residual work resolves the transfer of reference energy. The displacement virial adds the restoring-force projection required to describe a configurational response. When reference calculations are intermittent, their association with the evolving state becomes part of both quantities. These observations connect local force errors, reference allocation, and material dynamics within the same equations of motion.

The controlled tests establish complementary parts of this connection. Matching a complete reference timetable does not fix the energy transfer when that timetable is coupled to a different trajectory. In the anharmonic chain, an independent reference estimate predicts the configurational shift of a conservative base model despite its unchanged kinetic temperature. The reference estimates also rank candidate update rules by their expected thermal response. In silicon, the relative stiffness spectrum identifies a direction that scalar force calibration will damage, even as it improves mixed-configuration force errors. The subsequent nonlinear trajectories retain this predicted tradeoff and show the benefit of updating the correction along the actual motion. SnSe adds an independent finite-temperature configurational test: nonlinear reference energies and forces support the positive response anticipated from a separate calibration.

Sparse force records also retain useful directional information. Reanalysis of the lithium (Li) interface and tungsten (W) configurations from the preceding study gives finite-difference and finite-chord restoring-force discrepancies in larger systems (Appendix~\ref{app:prior-material-forces}). These records complement the new dynamical and configurational tests.

The comparison with established methods is consequential. The static-Hessian correction removes most of the silicon error, and the standard Strang scheme performs best in the present tests. The response framework supplies a common physical comparison: reference-process coefficients anticipate thermal and configurational shifts, while the relative stiffness spectrum identifies the directions that a correction will improve or damage. It complements advances in learned force constants, efficient Hessian evaluation, and learned multiple-time-step forces~\cite{phl2026,pft2026,dmts2026}. Such models can supply accurate and inexpensive directional information; the energy and virial response can then assess how that information is used during propagation.

The identities apply to a specified reference--approximation pair under the stated dynamical assumptions, while the demonstrations combine controlled model dynamics, nonlinear silicon trajectories, and local finite-temperature response integrals in SnSe. The stationary virial is one structural observable, and the silicon test concerns two collective coordinates over a fixed 240\,fs interval. For SnSe, the compact probe gives a well-defined local restoring-force response; sampling of the reference--surrogate difference currently limits its quantitative precision. Extending these calculations to diffusion, reactive events, or heat transport will require the response of those observables and the relevant sampling times. The held-force chain resolves a finite-mismatch limitation of the leading stationary-response prediction. Independently, the static-Hessian silicon trajectory shows that matching local curvature does not fully predict the response at finite amplitude. These results point to a practical next step: combine inexpensive local response information with occasional reference calculations targeted to the motion that matters, and judge the resulting dynamics against the physical observable being sought.

\appendix
\setcounter{table}{0}
\renewcommand{\thetable}{A\arabic{table}}
\renewcommand{\theHtable}{appendix.\arabic{table}}
\setcounter{figure}{0}
\renewcommand{\thefigure}{A\arabic{figure}}
\renewcommand{\theHfigure}{appendix.\arabic{figure}}

\section{Finite-window balances and response conditions}
\label{app:balances}
For a time window of length $L$, define the reference energy $E_{\rref}=K+U_{\rref}$, the moment $G=\bm q\cdot\bm p$, and $I=\bm q^{\mathsf T}\bm M\bm q/2$. Let $\mathcal P_L$ and $\mathcal V_L$ be Eqs.~\eqref{eq:power} and \eqref{eq:residual-virial} with expectations replaced by the corresponding finite-window integrals and event sums, divided by $L$. A consistent event convention includes the opening and closing half impulses in their propagation block. Direct application of It\^o's formula gives
\begin{align}
 \Delta E_{\rref}&=L\mathcal P_L-2\gamma L\overline K
                  +d\gamma\tau L+M_E,\label{eq:finite-energy}\\
 \Delta G&=L(2\overline K-\overline\Xi_{\rref}
                 +\mathcal V_L-\gamma\overline G)+M_G,\label{eq:finite-virial}\\
 \Delta I&=L\overline G.
\end{align}
Bars denote physical-time averages. The continuous martingales are
\begin{align}
 M_E&=\sqrt{2\gamma\tau}\int\bm v^{\mathsf T}\bm M^{1/2}\dd\bm W,\nonumber\\
 M_G&=\sqrt{2\gamma\tau}\int\bm q^{\mathsf T}\bm M^{1/2}\dd\bm W.
\end{align}
The exact jumps are $\Delta E_{\rref}=\bm v^-\cdot\Delta\bm p+\Delta\bm p^{\mathsf T}\bm M^{-1}\Delta\bm p/2$ and $\Delta G=\bm q\cdot\Delta\bm p$. There is no quadratic virial jump.

The finite-window structural balance can be written
\begin{align}
 \frac{\overline\Xi_{\rref}}{d\tau}-1
 ={}&\frac{\mathcal P_L}{d\gamma\tau}
    +\frac{\mathcal V_L}{d\tau}
    +\frac{M_E-\Delta E_{\rref}}{Ld\gamma\tau}\nonumber\\
   &+\frac{M_G-\Delta G-\gamma\Delta I}{Ld\tau}.
 \label{eq:finite-structure}
\end{align}
Thus finite-window endpoint terms need not vanish for individual trajectories. Equations~\eqref{eq:temperature} and \eqref{eq:structure} require a stationary extended process, integrable force and impulse contributions, mean-zero martingales, and endpoint expectations divided by $L$ tending to zero. The reference canonical density must be normalizable on the reduced coordinate space, with vanishing integration-by-parts boundary terms. A fixed-period correction uses its periodic steady state averaged over phase. These assumptions concern physical-time averages rather than boundary momentum samples.

For the small-perturbation prediction, continuity is required of the stationary \emph{marked} process, including the anchor and event rule. The relevant long-time and small-perturbation limits must be interchangeable, and the impulse second-moment rate must remain bounded. The explicit factor $\eps$ then leaves the reference averages in Eq.~\eqref{eq:reference-estimators} at leading order; the quadratic impulse energy enters at order $\eps^2$. These conditions justify an $o(\eps)$ remainder. A uniform $O(\eps^2)$ claim would require additional response smoothness.

For fixed-interval Strang, set $\bm g=\nabla V$ and $\phi=\bm q\cdot\bm g$. Each half impulse is $-\eps h\bm g/2$, whereas the continuous residual is $+\eps\bm g$. The leading virial contribution over one reference segment is
\begin{equation}
 \int_0^h\phi(\bm q_t)\dd t
 -\frac h2[\phi(\bm q_0)+\phi(\bm q_h)].
\end{equation}
Its stationary expectation is zero. The continuous power is the endpoint difference of $V$, with zero stationary mean, and the endpoint impulse-power means vanish by the canonical momentum symmetry. This proves the stated first-order cancellation for fixed $h$ and exact fast propagation. Numerical splitting and a state-dependent outer interval change the corresponding assumptions.

For coherent, unthermostatted material dynamics, $\gamma=0$ and there are no stochastic terms. Equations~\eqref{eq:finite-energy} and \eqref{eq:finite-virial} give $\Delta E_{\rref}=L\mathcal P_L$ and $\overline\Xi_{\rref}-2\overline K=\mathcal V_L-\Delta G/L$ for exact propagation. We retain the endpoint term and assess the silicon trajectories directly; stationary temperature formulas are not used for that experiment.

\section{Mode resolution and scalar covariance}
\label{app:modes}
For a harmonic reference, diagonalize the mass-weighted force constants,
\begin{equation}
 \bm M^{-1/2}\bm H_{\rref}\bm M^{-1/2}\bm u_j
 =\omega_j^2\bm u_j,
\end{equation}
and define $Q_j=\bm u_j^{\mathsf T}\bm M^{1/2}\bm q$ and $P_j=\bm u_j^{\mathsf T}\bm M^{-1/2}\bm p$. Projecting the continuous residual and impulses with $\bm u_j^{\mathsf T}\bm M^{-1/2}$ gives modal rates $\mathcal P_j$ and $\mathcal V_j$. Their stationary balances are
\begin{align}
 \langle P_j^2\rangle/\tau-1&=\mathcal P_j/(\gamma\tau),\\
 \omega_j^2\langle Q_j^2\rangle/\tau-1
 &=\mathcal P_j/(\gamma\tau)+\mathcal V_j/\tau.
\end{align}
No simultaneous diagonalization of the base force constants is needed if the complete projected residual is retained. For a reference with an anharmonic part $U_{\mathrm{anh}}$, the equations instead contain
\begin{align}
 \gamma(\langle P_j^2\rangle-\tau)
 &=\mathcal P_j-\langle P_j\partial_jU_{\mathrm{anh}}\rangle,\\
 \omega_j^2\langle Q_j^2\rangle-\langle P_j^2\rangle
 &=\mathcal V_j-\langle Q_j\partial_jU_{\mathrm{anh}}\rangle.
\end{align}
These terms carry anharmonic energy exchange and restoring forces. The silicon waveform coordinates use equal-mass Euclidean unit directions and lengths in \AA, as defined in Appendix~\ref{app:silicon}, rather than these mass-weighted canonical coordinates.

For positive harmonic force constants, the reference and scalar-corrected covariances are $\bm C_{\rref}=\tau\bm H_{\rref}^{-1}$ and $\bm C_c=\tau(c\bm H_{\bbase})^{-1}$. Consequently $\bm C_{\rref}^{-1/2}\bm C_c\bm C_{\rref}^{-1/2}=\bm G/c$. Every eigenvalue of $\bm G$ lies between $a$ and $b$, so the operator-norm error is $\max(|a/c-1|,|b/c-1|)$. Balancing the endpoint errors gives $c_*=(a+b)/2$ and Eq.~\eqref{eq:scalar-bound}. The bound concerns the worst-direction covariance error for any single scalar, including one fitted to a chosen mode. It neither bounds that mode's error alone nor constrains a matrix-valued or time-dependent correction.

A force-fit comparison uses the same matrices with two displacement weights. For a covariance $\bm\Sigma$, minimizing $\E\|(c\bm H_{\bbase}-\bm H_{\rref})\bm q\|^2$ gives
\begin{equation}
 c_F(\bm\Sigma)=
 \frac{\operatorname{Tr}(\bm H_{\bbase}\bm H_{\rref}\bm\Sigma)}
 {\operatorname{Tr}(\bm H_{\bbase}^2\bm\Sigma)}.
\end{equation}
The isotropic and reference-thermal choices are $\bm\Sigma\propto\bm I$ and $\bm\Sigma=\tau\bm H_{\rref}^{-1}$. They give $c_F=1.34797$ and $1.32532$, respectively. Table~\ref{tab:scalar-controls} evaluates them alongside the optical fit and covariance minimax. These matrix-based fits use more reference information than the earlier optical fit and optimize their specified force or covariance objective. The nonlinear trajectory comparison retains the original frozen scalar.

\begin{table*}[!tbp]
\caption{\label{tab:scalar-controls}Scalar corrections with different fitting objectives. Full-force errors use three saved density-functional configurations: two mixed displacements of atomic root-mean-square magnitude 0.01 and 0.02\,\AA, and one low-frequency displacement. Ratios and covariance errors use the 21-dimensional harmonic model. The optical fit predates these tests; the other fitted scalars use the complete force-constant matrices and do not use the three test-force labels.}
\centering
\begin{ruledtabular}
\begin{tabular}{lrrrrrrr}
 & & \multicolumn{3}{c}{Full-force error (\%)} & \multicolumn{2}{c}{Variance ratio} & Worst covariance\\
Objective & $c$ & Mixed 0.01 & Mixed 0.02 & Low & Low & Optical & error (\%)\\
\hline
Base & 1.00000 & 26.93 & 26.66 & 0.81 & 1.012 & 1.407 & 40.67\\
Optical fit & 1.37685 & 3.58 & 3.89 & 36.60 & 0.735 & 1.022 & 26.52\\
Covariance minimax & 1.20921 & 11.90 & 11.62 & 19.97 & 0.837 & 1.163 & 16.33\\
Isotropic force MSE & 1.34797 & 3.67 & 3.75 & 33.74 & 0.751 & 1.044 & 24.94\\
Thermal force MSE & 1.32532 & 4.51 & 4.43 & 31.49 & 0.763 & 1.061 & 23.66\\
\end{tabular}
\end{ruledtabular}
\end{table*}

\section{Model dynamics and statistical estimators}
\label{app:statistics}
\subsection{Harmonic matched-timetable test}
For $U_{\rref}=q^2/2$ and $U_{\bbase}=0.95q^2/2$, a held correction gives the linear equation $\dd v=[-0.95q-0.05a-\gamma v]\dd t+\sqrt{2\gamma\tau}\dd W$ within each anchor interval. We use its exact Gaussian transition, including the covariance of the integrated noise. The feedback event is tested after each completed monitoring interval and occurs if $|q-a|\ge0.2$ or the anchor age reaches 0.5. The replay member receives every feedback event, including during the discarded initial window, and refreshes its own anchor. Each member draws $q(0)$ and $v(0)$ independently from standard normal distributions, the unit reference canonical law, and uses independent driving noise. The 256 pairs are propagated to time 1200, with physical-time statistics accumulated over 200--1200. The finer monitoring grid is coupled through exact Gaussian noise composition. The replay is a controlled comparison requiring a donor trajectory, rather than a stand-alone low-cost scheduling algorithm.

\subsection{Anharmonic chains and reference clocks}
The periodic eight-particle chain has bond displacements $x_i=q_{i+1}-q_i$ and potentials
\begin{equation}
 U_2=\frac12\sum_{i=0}^{7}x_i^2,\qquad
 U_4=\frac12\sum_{i=0}^{7}x_i^4.
\end{equation}
The masses are $(1,2,1.5,1,2,1.5,1,2)$ in reduced units. Define $e_i=\sqrt{m_i/\sum m_i}$ and $\bm\Pi=\bm I-\bm e\bm e^{\mathsf T}$. Initial positions are zero and initial momenta are $\sqrt\tau\bm M^{1/2}\bm\Pi\bm\xi$, where $\bm\xi$ is a standard Gaussian vector. This fixes the center of mass and removes total momentum without rescaling the kinetic energy. The positions are initially nonequilibrium. All chain statistics use the fixed observation window 200--1200.

Dynamics use BAOAB~\cite{leimkuhler2013}: half force kick, half position drift, a full Ornstein--Uhlenbeck (OU) step, half drift, and a final half force kick. The two microsteps are 0.02 and 0.01. Hold stores the reference-minus-base force at its last anchor. Strang surrounds a fast BAOAB block with reference half impulses; the physical-time kinetic average uses endpoints after the opening and before the closing impulse. The instantaneous impulses carry no sampling duration.

For the anharmonic matched-timetable comparison, the feedback event uses the unweighted root-mean-square change of $\nabla U_2$ from its anchor, with threshold 0.2 or maximum age 0.5. This analytic reference-gradient trigger isolates feedback; its evaluation cost is not evidence for a deployable DFT savings. A separate reference trajectory accompanies each member for scoring, with shared initial state and noise but no feedback into the tested method. The reference prediction uses 128 calibration pairs. An initial 128-pair validation did not resolve the contrast sufficiently. The reported confirmation uses 256 new pairs and a noise-control estimator fixed before those runs, with the original prediction unchanged. Coarse and fine confirmation paths share coupled OU noise; the earlier and later validation data are kept separate.

For prospective selection, the candidate interval at anchor $n$ is
\begin{equation}
 h_n=\Delta t\,\operatorname{clip}\!\left(
 \left\lfloor\frac{C}{\Delta t\sqrt{2K_n/(d\tau)}}\right\rfloor,
 1,\operatorname{round}\frac{0.5}{\Delta t}\right).
 \label{eq:clock}
\end{equation}
Zero anchor kinetic energy gives the maximum interval. The 17 candidates have $C=0.180+0.005j$, $j=0,\ldots,16$. Among candidates whose calibration mean query rate is at most five, we minimize the absolute predicted temperature shift, breaking ties by fewer queries and then larger $C$. This yields $C=0.205$. Selection and total-virial validation use separate sets of 128 reference calibration paths and 256 validation paths per grid. Within a validation grid, methods share initial states and noise; calibration, coarse validation, and fine validation are mutually independent in these two experiments. Halving the microstep also changes the quantization of the selected clock in Eq.~\eqref{eq:clock}.

\subsection{Noise controls and uncertainty}
\begin{table*}[t]
\caption{\label{tab:virial-statistics}Reference displacement-virial shifts in the anharmonic chain. Values are percentages of $d\tau$. Each cell gives the paired method-minus-reference mean and its 95\% trajectory-bootstrap interval. CV denotes the fixed noise control in Eq.~\eqref{eq:control-variates}. The structural signal of the base force is resolved on both grids with CV; its raw half-step interval includes zero.}
\centering
\begin{ruledtabular}
\begin{tabular}{lrrrr}
 & \multicolumn{2}{c}{$\Delta t=0.02$} & \multicolumn{2}{c}{$\Delta t=0.01$}\\
Method & Raw & CV & Raw & CV\\
\hline
Base & $0.971$ $(0.032,1.933)$ & $0.652$ $(0.544,0.758)$ & $0.641$ $(-0.201,1.547)$ & $0.537$ $(0.413,0.663)$\\
Hold & $4.531$ $(3.500,5.533)$ & $3.512$ $(3.390,3.631)$ & $3.926$ $(2.971,4.890)$ & $3.476$ $(3.337,3.613)$\\
Strang & $0.159$ $(-0.646,0.923)$ & $-0.024$ $(-0.102,0.053)$ & $-0.330$ $(-1.089,0.418)$ & $-0.022$ $(-0.118,0.074)$\\
\end{tabular}
\end{ruledtabular}
\end{table*}

At an OU step, let $a_O=e^{-\gamma\Delta t}$, $b_O=\sqrt{\tau(1-a_O^2)}$, and $\bm z=\bm\Pi\bm\xi$. The momentum update is
\begin{equation}
 \bm p^+=a_O\bm p^-+b_O\bm M^{1/2}\bm z.
\end{equation}
The kinetic heat $Q_O=K^+-K^-$ has conditional mean $(a_O^2-1)K^-+(1-a_O^2)d\tau/2$. Define the centered increments
\begin{align}
 m_E&=Q_O-\E[Q_O\mid\text{pre-OU state}],\nonumber\\
 m_G&=\bm q_O\cdot(\bm p^+-a_O\bm p^-).
\end{align}
Both have conditional expectation zero. Using $L=1000$, we form
\begin{align}
 \widehat\theta_K^{\mathrm{CV}}
 &=\widehat\theta_K^{\mathrm{raw}}-\frac{\sum m_E}{Ld\gamma\tau},\nonumber\\
 \widehat\theta_\Xi^{\mathrm{CV}}
 &=\widehat\theta_\Xi^{\mathrm{raw}}-\frac{\sum m_E}{Ld\gamma\tau}
                       -\frac{\sum m_G}{Ld\tau}.
 \label{eq:control-variates}
\end{align}
The denominators are 87.5 and 1750. The coefficients are fixed, not fitted to outcomes. This subtraction preserves the finite-step raw estimator's expectation; it does not remove initialization or integration bias. Raw observables are computed directly from positions and momenta, without residual work or endpoint energy changes. Table~\ref{tab:virial-statistics} compares the two estimators.

The reported shifts are paired method-minus-reference averages on the same numerical grid. Each independent trajectory, or donor--receiver pair, is one resampling unit. We use 2000 bootstrap replicates and percentile 95\% intervals, preserving pairing among methods and the covariance between power and virial coefficients. Prediction-error intervals resample calibration and validation ensembles independently. The two grids are independently resampled in the selection and total-virial tests, and paired in the matched-timetable confirmation. Confidence intervals describe sampling uncertainty of the specified finite-window means. They do not count time steps as independent samples or establish equivalence from a zero-containing interval.

\begin{table}[!tbp]
\caption{\label{tab:response-checks}Statistical response checks. The upper block gives differences in normalized reference displacement-virial shifts, in percentage points (pp), using the fixed noise control. The base contrast is fine minus coarse; the hold errors are observation minus prediction. The lower block gives harmonic-oscillator residual second moments at monitoring interval 0.02, in the reduced units of Sec.~\ref{sec:statistics}. All intervals are 95\% trajectory-bootstrap intervals.}
\centering
\begin{ruledtabular}
\begin{tabular}{lrr}
Quantity & Mean & 95\% interval\\
\hline
\multicolumn{3}{c}{Displacement-virial differences (pp)}\\
Base, fine minus coarse & $-0.115$ & $(-0.287,0.053)$\\
Hold error, $\Delta t=0.02$ & $0.235$ & $(0.102,0.372)$\\
Hold error, $\Delta t=0.01$ & $0.199$ & $(0.049,0.349)$\\
\hline
\multicolumn{3}{c}{Residual second moment $\langle R^2\rangle$ ($10^{-5}$)}\\
Feedback & $3.00477$ & $(2.99000,3.01893)$\\
Replay & $12.76952$ & $(12.38307,13.15754)$\\
\end{tabular}
\end{ruledtabular}
\smallskip
\raggedright\footnotesize
The base contrast uses independent grid ensembles; hold errors use independent calibration and validation ensembles. Their intervals resolve the hold underprediction, while the base cross-grid difference remains unresolved. Residual second moments are physical-time averages over 200--1200 for 256 feedback--replay pairs. They quantify force amplitude alongside the signed residual work that governs heating.
\end{table}

\section{Electronic-structure and nonlinear silicon calculations}
\label{app:silicon}
\subsection{Reference, potential, and local force constants}
The fixed cubic eight-atom silicon cell has lattice parameter $5.431$\,\AA\ and atomic mass $28.085$\,u. Reference forces and energies are computed with Quantum ESPRESSO 7.5~\cite{qe2017}, PBE~\cite{pbe1996}, and D3 dispersion~\cite{d3_2010}. We use a PAW silicon pseudopotential, \texttt{Si.pbe-n-kjpaw\_psl.1.0.0.UPF}, plane-wave and density cutoffs of 50 and 400\,Ry, an unshifted $4\times4\times4$ $k$-point grid, fixed occupations, and a non-spin-polarized calculation. The self-consistent threshold is $10^{-8}$\,Ry; an initial-state repeat at $10^{-10}$\,Ry changes the force by 0.0055\% in relative norm. The base model is the medium MACE-MPA-0 potential~\cite{macefoundation2025}, evaluated in double precision, with its additional dispersion option disabled and no retraining.

Central force differences at a symmetry-independent atomic displacement of $\pm0.01$\,\AA\ are completed using the 192 crystal symmetry operations. The symmetric force-constant matrices are projected away from the three translations, leaving 21 positive modes. A $\pm0.02$\,\AA\ check changes the reference, base, and difference matrices by 0.042\%, 0.116\%, and 0.387\%, respectively, in relative Frobenius norm. The degeneracies of the four reference frequency groups are 6, 6, 6, and 3. The scalar multiplier $1.3768451271$ is inherited from an independent optical-direction force calibration. No nonlinear-trajectory data enter that fit.

\subsection{Initial state, propagation, and waveform objective}
Let $\bm u_{\mathrm{low}}$ be the Euclidean unit vector obtained by projecting atom 0's Cartesian $x$ direction into the lowest six-dimensional mode subspace, with its sign fixed by that component. Let $\bm u_{\mathrm{opt}}$ have alternating $x$ components $\pm1/\sqrt8$ on the two diamond sublattices and zero $y,z$ components. The directions are orthogonal. The initial displacement and velocity are
\begin{equation}
 \bm q(0)=\sqrt8(0.08\bm u_{\mathrm{low}}+0.02\bm u_{\mathrm{opt}})\,\text{\AA},
 \qquad \bm v(0)=0.
\end{equation}
All 24 Cartesian coordinates then evolve with the full nonlinear forces, without a thermostat or post-step center-of-mass correction. The motion is not restricted to the two plotted directions.

Reference trajectories use velocity Verlet at 1 and 2\,fs. The base, scalar, and static-Hessian trajectories use 0.5\,fs and independent 0.25\,fs checks. For Strang, each 4\,fs block consists of a half correction impulse, eight 0.5\,fs base-force Verlet steps, and a half correction impulse at the actual endpoint. The check uses sixteen 0.25\,fs inner steps with the same outer interval. Both half impulses have the correction sign $\bm F_{\rref}-\bm F_{\bbase}$.

Projected coordinates are $Q_j(t)=\bm u_j\cdot\bm q(t)$, with initial amplitudes $A_{\mathrm{low}}=0.226274$\,\AA\ and $A_{\mathrm{opt}}=0.056569$\,\AA. On the common 2\,fs output grid, trapezoidal quadrature over $T=240$\,fs gives
\begin{align}
 E_j^2&=\frac1T\int_0^T
       \left[\frac{Q_j^{\mathrm{method}}(t)-Q_j^{\mathrm{ref}}(t)}{A_j}\right]^2\dd t,\nonumber\\
 E_{\mathrm{joint}}&=\sqrt{(E_{\mathrm{low}}^2+E_{\mathrm{opt}}^2)/2}.
 \label{eq:waveform}
\end{align}
The two coordinates have equal weights in this specified objective. Their errors are deterministic waveform differences, not estimates from independent time samples. Harmonic forecasts use the same initial state and fixed force constants. The static-Hessian force equals the reference harmonic force in that limit. Its small nonzero harmonic waveform error arises from the different integration steps used for the candidate and reference trajectories. The static-Hessian comparator and its 2\% criterion were specified after the three-point initial-state pilot and before the complete trajectory results became available. Table~\ref{tab:si-errors} collects the predictions and numerical checks.

\begin{table}[!tbp]
\caption{\label{tab:si-errors}Joint waveform errors in silicon, in percent. The harmonic forecast was fixed before the nonlinear trajectories. The last column compares each main candidate with its own half-step trajectory; it is a numerical check, not an error against the DFT reference.}
\centering
\begin{ruledtabular}
\begin{tabular}{lrrrr}
 & Harmonic & \multicolumn{2}{c}{DFT reference} & Inner-step\\
Method & forecast & 1\,fs & 2\,fs & change\\
\hline
Base & 76.612 & 75.052 & 75.225 & 0.028\\
Scalar & 27.439 & 28.215 & 28.329 & 0.046\\
Hessian & 0.194 & 3.134 & 2.488 & 0.048\\
Strang & 0.250 & 0.217 & 0.674 & 0.044\\
\end{tabular}
\end{ruledtabular}
\smallskip
\raggedright\footnotesize
The two DFT trajectories differ by 0.760\%. The joint target is 2\%. Half-step candidate errors against the 1\,fs reference are 75.060\%, 28.221\%, 3.091\%, and 0.198\%, respectively.
\end{table}

\subsection{Reference accounting}
The whole-cell initial excitation energy is $0.09534475$\,eV above the equilibrium reference geometry. The 1 and 2\,fs reference trajectories contain 241 and 121 energy samples; their sampled peak-to-peak reference-energy variations are 0.123\% and 0.459\% of that excitation. Strang has 61 reference evaluations at outer boundaries. Base, scalar, and static-Hessian production branches have reference energies only at initialization and the final configuration; their half-step checks have only the common initial reference value. Their densely sampled approximate-potential energies are not dense reference-energy curves.

Deployment counts include the initial reference force. The material validation required 486 new self-consistent calculations in total: three initial-state checks, 240 and 120 new reference-trajectory evaluations, 60 Strang and 60 half-step Strang evaluations, and three terminal scoring evaluations. All converged, with no retries. These jobs used 86.97 allocated CPU core-hours. The earlier scalar calibration and force-constant study used ten additional self-consistent calculations and 5.67 allocated core-hours. Calibration, validation, and deployment costs are separate; the 61-versus-121 comparison concerns reference evaluations for propagation, rather than a measured end-to-end wall-time speedup.

\section{Independent crystal-response integrals}
\label{app:snse}
\subsection{Tin selenide reference and fixed probe}
The reference is an eight-atom $Pnma$ SnSe cell with the lattice parameters given in Sec.~\ref{sec:snse}. Internal coordinates are stationary at a maximum atomic force of $3.82\times10^{-5}$\,eV\,\AA$^{-1}$. All energies and forces use the actual nonlinear potentials at the sampled configurations. The reference calculation uses Quantum ESPRESSO~\cite{qe2017}, PBE with D3 dispersion~\cite{pbe1996,d3_2010}, scalar-relativistic PAW datasets and fixed occupations, plane-wave cutoffs of 100 and 800\,Ry for wavefunctions and charge density, a $5\times12\times12$ $k$ mesh, and an SCF convergence threshold of $10^{-10}$\,Ry. The base model is MACE-MPA-0 medium in double precision with added dispersion disabled. The reference--model difference therefore includes the specified electronic-structure target, as in the silicon comparison.

A fixed orthonormal matrix maps 21 translation-free coordinates to 24 Cartesian displacements. The reference and base force constants are positive on this subspace. Set $s=\bm q^{\mathsf T}\bm H_{\rref}\bm q/\tau$, with $\tau=0.0407599863$\,eV. For inner and outer squared radii $s_i$ and $s_o$, the probe is $\bm h=\bm q\chi(s)$, where $\chi=1$ below $s_i$, $\chi=0$ above $s_o$, and
\begin{equation}
 \chi=1-10t^3+15t^4-6t^5,\qquad
 t=(s-s_i)/(s_o-s_i)
\end{equation}
inside the transition layer. Its full divergence is $D_h=21\chi+2s\chi'$, including the signed transition contribution. The primary radii are $(20,40)$. Multiplying both squared radii by 0.9 and 1.1 defines two sensitivity probes before validation. The common support lies inside a single periodic displacement chart. Equation~\eqref{eq:crystal-probe} is a local generalized reference-virial ratio; it does not require the probe region to exhaust the bulk equilibrium distribution.

\subsection{Independent prediction and observation}
A fixed Gaussian-mixture proposal samples configuration pairs. In coordinates whitened by $\bm H_{\rref}/\tau$, it has a 0.2 standard-normal component and a 0.8 shifted Gaussian component fitted before the validation draw. One component is selected independently for each pair, followed by opposite deviations about its mean. Each marginal follows the full mixture density $\psi$. The factor $\rho=\phi/\psi$, where $\phi=N(0,\tau\bm H_{\rref}^{-1})$, includes Gaussian determinants and satisfies $\rho\le5$. The configuration pair, rather than each member or each force component, is the independent sampling unit.

Write $\Delta E_j(\bm q)=U_j(\bm q)-U_j(0)$ and $E_H=\bm q^{\mathsf T}\bm H_{\rref}\bm q/2$. On the probe support, define
\begin{equation}
 w_\lambda=\rho\exp\!\left[-\frac{(1-\lambda)\Delta E_{\rref}
                   +\lambda\Delta E_{\bbase}-E_H}{\tau}\right].
\end{equation}
The response observation uses the known zero-order identity $I_0[X_h-D_h]=0$. It is estimated through the integrands
\begin{align}
 k_\lambda&=(w_\lambda-w_0)(X_h-D_h),\nonumber\\
 b_\lambda&=w_\lambda D_h-\rho(D_h-\mu_D),
 \label{eq:probe-estimator}
\end{align}
where $\mu_D=\E_\phi D_h$ is obtained by one-dimensional Gaussian radial integration. For the primary probe $\mu_D=17.250812812$. The second term in $b_\lambda$ has known mean zero. Target terms vanish outside the support, but centered Gaussian controls there are retained. The ratio is $\widehat\theta_{h,\lambda}=1+\widehat I[k_\lambda]/\widehat I[b_\lambda]$. The reference-level response integrand uses reference forces and reference/base energies, with no base force or fitted $\beta_{h,0}$. The controlled value at zero perturbation is exactly one by construction. As a separate baseline diagnostic, the uncontrolled zero-perturbation ratio is 0.9686 with standard error 0.0374.

To reduce the reference cost, a surrogate corrects the base potential's gradient and Hessian at the center,
\begin{equation}
 U_s=U_{\bbase}+\bm q\cdot\Delta\bm g
       +\tfrac12\bm q^{\mathsf T}(\bm H_{\rref}-\bm H_{\bbase})\bm q,
\end{equation}
where $\Delta\bm g=\nabla U_{\rref}(0)-\nabla U_{\bbase}(0)$. The surrogate integrands use $\bm F_s=-\nabla U_s$, which includes the base force and the fixed gradient and Hessian corrections. Each required integral vector is estimated as a model mean plus a paired reference-minus-surrogate correction,
\begin{equation}
 \widehat{\bm I}=\overline{\bm u}_M+\overline{\bm v}_n,
 \qquad
 \widehat{\operatorname{Cov}}(\widehat{\bm I})
 =\frac{\bm S_u}{M}+\frac{\bm S_v}{n}.
 \label{eq:two-level-probe}
\end{equation}
Here $\bm u$ contains surrogate integrands and $\bm v$ is the reference-minus-surrogate difference at the same configuration pair. The two means use independent draws, $M=16\,384$ and $n=64$. All 128 reference labels are retained, including the cutoff layer. A separate calibration uses 32 reference pairs and 8192 independent model pairs. It supplies the frozen coefficient 0.06585894 and standard error 0.00446456, including both variance layers.

\subsection{Uncertainty and numerical sensitivity}
\begin{table}[!tbp]
\caption{\label{tab:snse}SnSe probe sensitivity. Entries are shifts $100(\theta_{h,\lambda}-1)$ with approximate 95\% Fieller intervals, in percent. Each scale multiplies both squared cutoff radii; all entries use the same 64 independent reference pairs and 16\,384 independent model pairs. The controlled zero-perturbation value is fixed at zero shift.}
\centering
\begin{ruledtabular}
\begin{tabular}{cccc}
Scale & $\lambda=0.25$ & $\lambda=0.5$ & $\lambda=1$\\
\hline
0.9 & 2.36 & 4.63 & 8.39\\
 & (1.25, 3.52) & (2.82, 6.53) & (4.34, 12.46)\\
1.0 & 2.63 & 5.09 & 9.84\\
 & (0.99, 4.33) & (2.56, 7.70) & (4.11, 15.60)\\
1.1 & 2.24 & 4.68 & 12.29\\
 & (0.75, 3.85) & (2.22, 7.30) & (6.21, 18.45)\\
\end{tabular}
\end{ruledtabular}
\end{table}

Fieller intervals retain the covariance between the response increment and denominator in Eq.~\eqref{eq:two-level-probe}, using a two-sided 95\% $t_{63}$ critical value. Observation-minus-prediction intervals instead use the delta-method variance of the response ratio plus the independent variance of $\lambda\beta_{h,0}$. We construct these symmetric intervals with $t_{31}$. Both are approximate working intervals. The primary denominator has relative standard error 3.22\% and a positive lower confidence bound. One pair contributes 67.8\% of the estimated response variance, so the interval precision remains sensitive to the sampled reference--surrogate tail. Full pair deletion diagnostics retain positive shifts from 1.940\% to 2.785\%; they are sensitivity calculations, not grounds for excluding configurations.

The response interval has a half-width of 1.67 percentage points. Including calibration uncertainty, the observed-minus-predicted shift has an interval of $[-0.73,2.70]$ percentage points, with half-width 1.72. These absolute uncertainties refer to the normalized virial ratio. The reference--surrogate difference contributes 97.2\% of the estimated response variance, identifying the main source of sampling uncertainty.

The most influential pair was recomputed at 120/960\,Ry, a $6\times14\times14$ $k$ mesh, and an SCF threshold of $10^{-11}$\,Ry. Subtracting the corresponding strict center energy changes the two energy increments by 0.146 and 0.191\,meV, and the force vectors by 0.0203\% and 0.0513\% in relative norm. The largest logarithmic-weight change is 0.00469. Replacing this pair in a sensitivity copy gives a shift of 2.638\%, an interval of $[0.983,4.342]\%$, and a change of 0.0053 percentage points. This isolates the impact of tighter electronic settings at the dominant pair; it is not a uniform convergence bound for all configurations.

A separately fixed smaller probe, $(s_i,s_o)=(10,20)$, gives a 2.041\% shift with interval $[0.383,3.399]\%$. Its frozen prediction is 1.271\%. The prediction-error interval is $[-0.862,2.403]$ percentage points. The interval is narrower, while the largest pair contributes 74.4\% of the estimated response variance. The original probe remains the primary material comparison.

\subsection{Equivalent response expressions}
\label{app:estimator-equivalence}
Equation~\eqref{eq:crystal-probe} provides a comparison between two ways to measure the response at the same perturbation. The first uses reference forces and energy differences in Eq.~\eqref{eq:probe-estimator}; the second evaluates the residual-force integral on the right-hand side of Eq.~\eqref{eq:crystal-probe}. Their difference has target zero. We evaluate it with the same 64 reference pairs and model pool, retaining the covariance of both numerators and their common denominator. At $\lambda=0.25$, the difference between the two shifts is 1.018 percentage points, with standard error 0.737 and an approximate 95\% interval of $[-0.455,2.491]$ percentage points. The corresponding intervals are $[-0.603,3.838]$ at $\lambda=0.5$ and $[-1.699,5.794]$ at $\lambda=1$. These comparisons test internal consistency of the two estimators; the frozen prediction remains the one from the separate calibration.

The zero-perturbation derivatives can also be compared directly. Define $\delta=(\Delta E_{\bbase}-\Delta E_{\rref})/\tau$. Differentiating the controlled numerator gives
\begin{equation}
 \left.\partial_\lambda k_\lambda\right|_0
 =-w_0\delta(X_h-D_h).
 \label{eq:energy-response-derivative}
\end{equation}
On the validation sample, its ratio to the reference denominator is $0.1098\pm0.0424$, whereas the residual expression for $\beta_{h,0}$ gives $0.05947\pm0.00692$; errors here are standard errors. The paired difference is $0.0504\pm0.0377$, with an approximate 95\% interval of $[-0.0250,0.1257]$. The standard error of the energy-response derivative is 6.1 times that of the residual-force coefficient. The latter differs from the separate 32-pair calibration by 0.78 combined standard errors. The derivative and all three finite-perturbation observations are evaluated from the same sampled configurations and share sampling fluctuations. Their covariance is relevant to interpreting departures from the leading forecast alongside nonlinear response.

\subsection{Weight and influence diagnostics}
\label{app:weight-diagnostics}
For each reference pair let $\bar w_j$ be the mean of its two positive importance weights. The descriptive weight count $(\sum_j\bar w_j)^2/\sum_j\bar w_j^2$ is 25.20, 24.17, 21.70, and 16.32 for $\lambda=0$, 0.25, 0.5, and 1, respectively, out of 64 independent pairs. At the primary perturbation the largest pair carries 14.60\% of the total positive weight. The response precision additionally depends on the signed probe divergence, the reference--surrogate difference, and their covariance. Accordingly, the largest pair's 67.8\% contribution to the estimated response variance is a separate, more concentrated diagnostic; the weight count is not an effective sample size for the final controlled ratio.

The compact probe also separates moment existence from finite-sample precision. If the reference and model energies and forces are continuous on the compact support and the support contains no potential singularity, they are bounded there. With $\rho\le5$, the target weights and probe integrands are then bounded. The Gaussian controls remain bounded outside the support as well. These conditions give finite moments for the integral estimators. They do not bound the error of a particular 64-pair realization; the observed concentration of the reference--surrogate variance remains relevant to the working intervals.

\subsection{Nonzero responses with known answers}
\label{app:known-answer-response}
\begin{table}[!htbp]
\caption{Known nonzero responses and coverage of nominal 95\% Fieller intervals, using $t_{63}$ and 250 replicates for each example. H$+$ and H$-$ denote the harmonic cases with $\eps=+0.2$ and $-0.2$; A denotes the anharmonic example. The binomial Monte Carlo standard errors of the coverage fractions are 1.11--1.40 percentage points. Different perturbations within an example use correlated samples.}
\label{tab:known-answer-response}
\begin{ruledtabular}
\begin{tabular}{lrrrr}
Example & $\lambda$ & Exact/quad. $\theta$ & Mean $\widehat\theta$ & Covered\\
H$+$ & 0.25 & 1.052632 & 1.052430 & 242/250\\
     & 0.50 & 1.111111 & 1.110685 & 242/250\\
     & 1.00 & 1.250000 & 1.249048 & 241/250\\
H$-$ & 0.25 & 0.952381 & 0.952504 & 237/250\\
     & 0.50 & 0.909091 & 0.909325 & 237/250\\
     & 1.00 & 0.833333 & 0.833759 & 239/250\\
A    & 0.25 & 1.102409 & 1.102580 & 240/250\\
     & 0.50 & 1.244609 & 1.245498 & 239/250\\
     & 1.00 & 1.829853 & 1.839599 & 239/250\\
\end{tabular}
\end{ruledtabular}
\end{table}
We tested the integrands in Eq.~\eqref{eq:probe-estimator} with independently known nonzero responses. The potentials, proposal, seeds, and sample sizes were fixed before simulation. Each of 250 independent replicates used $n=64$ reference pairs and $M=1024$ model pairs, with the two-level covariance in Eq.~\eqref{eq:two-level-probe}. The pair construction and full mixture density follow the preceding estimator, including the signed cutoff derivative and Gaussian controls outside the support. Different $\lambda$ values shared samples within each replicate.

For the three-dimensional harmonic checks, $U_{\rref}=\bm q^{\mathsf T}\bm H\bm q/2$ and $U_{\bbase}=(1-\eps)U_{\rref}$, with $\eps=\pm0.2$. The compact-probe identity gives $\theta_{h,\lambda}=(1-\lambda\eps)^{-1}$ and $\beta_{h,0}=\eps$. We used $\tau=0.7$, a nondiagonal positive-definite $\bm H$, and squared cutoff radii $(4,9)$. Here the corrected surrogate equals the reference, so the difference layer has only roundoff-level variance.

To test a nonzero difference layer, we also used a one-dimensional anharmonic example, with $x=\sqrt{H/\tau}\,q$, $H=1.7$, $\tau=0.7$, and squared radii $(1,9)$:
\begin{align}
 U_{\rref}/\tau&=\tfrac12x^2+0.08x^3+0.12x^4,\nonumber\\
 U_{\bbase}/\tau&=0.4x^2-0.04x^3+0.04x^4,\nonumber\\
 U_s/\tau&=\tfrac12x^2-0.04x^3+0.04x^4.
\end{align}
Independent adaptive quadrature of the raw virial numerator and signed divergence denominator supplied the reference answers. They agreed with separate quadrature of the controlled integrands to below $10^{-12}$ in the ratio. The largest propagated quadrature error estimate was $6.4\times10^{-12}$. No response coefficient enters the observation estimator.

Table~\ref{tab:known-answer-response} gives the results. Coverage ranges from 94.8\% to 96.8\%, with no failed replicates or unbounded or empty intervals among the 2250 nonzero evaluations. Both variance layers were nonzero in every anharmonic replicate. Their mean difference-layer shares were 94.87\%, 96.33\%, and 98.90\% at the three nonzero perturbations. At $\lambda=1$, the anharmonic estimate had a bias of $0.00975$ with Monte Carlo standard error $0.00682$. Its empirical standard deviation was $0.10784$, compared with a root-mean-square estimated standard error of $0.11273$. These low-dimensional benchmarks test the estimator and its working intervals; SnSe precision is assessed separately from the material samples.

\section{Directional force discrepancies in extended systems}
\label{app:prior-material-forces}

The Li-interface and tungsten force data of Ref.~\cite{kang2026} provide complementary illustrations of the restoring-force information carried by a residual. On a fixed local displacement branch, let $\bm x=\bm q-\bm a$ and $\bm c=\bm F_{\rref}(\bm a)-\bm F_{\bbase}(\bm a)$. For conservative forces, the residual $\bm R=\bm F_{\bbase}+\bm c-\bm F_{\rref}$ satisfies $\nabla_{\bm q}W_{\bm a\to\bm q}=\bm R$, connecting Eq.~\eqref{eq:segment-work} to the pointwise projection $\bm x\cdot\bm R$. A quadratic residual potential gives $\bm x\cdot\bm R=2W$; this simplification does not hold generally.

For the 474-atom Li interface, paired probes along two unit configuration-space directions at each of two origins measure
\begin{equation}
 k_{\alpha}(\delta)=-\bm e\cdot
 \frac{\bm F_{\alpha}(\bm a+\delta\bm e)-\bm F_{\alpha}(\bm a-\delta\bm e)}{2\delta},
 \qquad \alpha\in\{\rref,\bbase\}.
\end{equation}
At $\delta=0.02$\,\AA, the four ratios $k_{\rref}/k_{\bbase}$ are 1.01689, 1.01084, 1.01987, and 1.01978. Doubling the full-configuration displacement amplitude to $0.04$\,\AA\ gives 1.01688, 1.01087, 1.01976, and 1.01978. The base restoring force is therefore slightly softer in all four sampled directions; a constant origin correction leaves these finite differences unchanged.

At a diagnostic 1\,fs endpoint of one fixed-correction interface trajectory, halving the timestep from 0.125 to 0.0625\,fs changes $\bm x\cdot\bm R$ from 0.07798720 to 0.07798152\,eV, whereas $W$ changes from 0.03988956 to 0.03988084\,eV. This comparison supports the numerical stability of that finite-endpoint projection, beyond the trajectory's previously admitted prefix.

For 432-atom tungsten, the chord from the origin to a previously identified 50\,fs spike configuration has atomic root-mean-square displacements of 0.390805\,\AA\ in the core and 0.074529\,\AA\ in the matrix. Defining $k_{\alpha}^{\mathrm{sec}}=-\bm x\cdot\Delta\bm F_{\alpha}/(\bm x\cdot\bm x)$, the reference-to-base ratios are 1.13249, 1.13329, and 1.12991 for the perturbed committee, stock committee, and stock committee with additive D3 forces, respectively. All three comparisons use the same two configurations, not separate policy trajectories.

These directional derivatives and finite-chord projections characterize force discrepancies relative to the specified density-functional references. They establish neither a full local covariance spectrum nor an independently measured thermal response. In particular, the unthermostatted, four-femtosecond interface paths cannot supply the stationary averages in Eqs.~\eqref{eq:temperature} and~\eqref{eq:structure}.

\subsection*{Computational assistance}
The authors directed the study and were responsible for its scientific conception, computational methods, validation design, and interpretation. Large language model tools assisted code development and debugging, data organization and analysis, checks of derivations, and manuscript preparation and language revision. The authors reviewed the assisted work and take responsibility for the final manuscript.

\subsection*{Author contributions}
P.~K.\ conceived and designed the study and developed its theoretical framework and computational protocols. He led the calculations, validation, and analysis and wrote the manuscript. L.~Z.\ and L.-D.~Z.\ secured funding and provided scientific guidance. D.~W., S.~B., P.~Z., P.~W., C.~W., Z.~L., and Y.~L.\ contributed to data preparation and selected validation calculations.

\bibliography{references}
\end{document}